\documentclass[aps,pre,preprint,notitlepage,longbibliography,superscriptaddress, showkeys]{revtex4-1}

\usepackage{amssymb}
\usepackage{amsmath}
\usepackage{mathtools}
\usepackage{graphicx}
\usepackage[]{natbib}
\usepackage{color}
\usepackage[pageanchor=true,
            plainpages=false,
            pdfpagelabels,
            bookmarks,
            bookmarksnumbered,
            colorlinks=true,
            allcolors=blue,
            pdfstartview=FitH]{hyperref}

\newcommand{\rp}{{\tt rigidPy}}
\newcommand{\note}[1]{{\color{black}{#1}}}

\newcommand\bondsign{\mathrel{\bullet\mkern-3mu{-}\mkern-3mu\bullet}}

\begin{document}

\title{rigidPy: Rigidity Analysis in Python}
\author{Varda F. Hagh}
\email[]{Corresponding author: vardahagh@uchicago.edu}
\affiliation{James Franck Institute, University of Chicago, Chicago, IL 60637, USA}
\title{rigidPy: Rigidity Analysis in Python}
\author{Mahdi Sadjadi}
\email[]{mahdisadjadi@asu.edu}
\affiliation{Department of Physics,
Arizona State University, Tempe, AZ 85287-1604}

\begin{abstract}
\rp{} is a Python package that provides a set of tools necessary for studying rigidity and mechanical response in elastic networks. It also includes suitable modules for generating new realizations of networks with applications in glassy systems and protein structures. \rp{} is available freely on \texttt{GitHub} and can be installed using Python Package Index (PyPi). The detailed setup information is provided in this paper, along with an overview of the mathematical framework that has been used in developing the package. 

\end{abstract}

\maketitle

{\bf PROGRAM SUMMARY}
\begin{small}
\noindent \\
{\em Program Title:} \rp{}\\
{\em Journal Reference:} \\
{\em Catalogue identifier:}\\
{\em Licensing provisions:} MIT license (MIT)  \\
{\em Developer's repository link:} \href{github.com/vardahagh/rigidpy}{github.com/vardahagh/rigidpy} \\
{\em Programming language:} Python $>3.5$ \\
{\em Nature of problem:} Elasticity, Linear Response, and Structural Optimization \\
{\em Solution method:} Linear Algebra, Convex Optimization
\end{small}
\newpage
\section{Introduction}
\label{sec:intro}
Rigidity theory involves the study of structural stability in mechanical systems as a result of the balance between their degrees of freedom and constraints. Its applications span many areas of research from structural engineering~\cite{calladine1978buckminster,connelly1996second,kato1994effect,grigorjeva2010static}, robotics~\cite{krick2009stabilisation, steltz2010jamming, zelazo2012rigidity}, and wireless network localization~\cite{aspnes2006theory,zhao2018bearing} to \note{studies of amorphous solids~\cite{he1985elastic, micoulaut2007onset, vaagberg2011glassiness, lopez2013jamming, Ellenbroek2015, lubensky2015phonons,charbonneau2016universal}} and biological systems such as confluent tissues~\cite{bi2015density,bi2016motility,merkel2018geometrically,yan2019multicellular}, biopolymer networks~\cite{storm2005nonlinear,huisman2011internal,rens2018micromechanical}, and proteins~\cite{jacobs2001protein,perticaroli2013secondary,perticaroli2014rigidity,karshikoff2015rigidity,atilgan2001anisotropy,doruker2000dynamics}.
First attempts in formulating a mathematical description of rigidity date back to Maxwell who studied principles of constructing stiff bar-and-joint frames~\cite{maxwell1864calculation}. Maxwell considered a frame as a set of joints that are connected via rigid bars. A mechanical frame of $N$ joints in $d=3$ dimensions has $3N$ degrees of freedom since each joint has three \note{translational} degrees of freedom. Maxwell realized that to render the set of $3N$ joints rigid, $3N - 6$ bars (constraints) are required where subtracting $6$ accounts for the trivial rigid motions including $d=3$ translations and $d(d-1)/2 = 3$ rotations if the frame has free boundary conditions. This counting rule would be different if the frame has periodic boundary conditions since a periodic structure does not have rotational degrees of freedom. Calladine, later, modified Maxwell's rule to take into account the existence of redundant bars that lead to states of self-stress~\cite{calladine1978buckminster}. In the Maxwell-Calladine count, the difference between the number of degrees of freedom ($Nd$) and constraints ($N_c$) in a $d$-dimensional frame is equal to the difference between its number of floppy modes ($F$) and the number of states of self-stress ($S$)\note{\cite{lubensky2015phonons}}:
\begin{equation}
F - S = Nd - N_c.
\label{eq:maxwell-calladine}
\end{equation}
A system with no states of self-stress and no non-trivial floppy modes is called \note{\textit{isostatic}}. Eq.~(\ref{eq:maxwell-calladine}) can be used to describe the rigidity of bar-and-joint structures or other physical systems that become rigid when there are enough constraints to cancel out the existing degrees of freedom. Such systems are called first-order rigid. However, this equation is not a suitable proxy for measuring rigidity in systems with higher-order rigidity such as under-constrained spring networks that rigidify under tension~\cite{damavandi2021energetic,damavandi2021energetic2}. For this reason, it is more appropriate to characterize the rigidity of a system using changes in its energy due to infinitesimal deformations. These deformations can be in the form of shear, hydrostatic pressure, or displacements of individual particles~\cite{schlegel2016local,zaccone2011approximate}. 

In the following sections, we first present a mathematical framework that provides a robust proxy for measuring rigidity of physical systems~\cite{damavandi2021energetic}, and then we demonstrate how the introduced tools can be computed in \rp{} for Hookean spring networks of arbitrary size. 

 
 \section{Mathematical Background}\label{sec:theory}

\begin{figure}[t]
\centering
\includegraphics[scale=0.4]{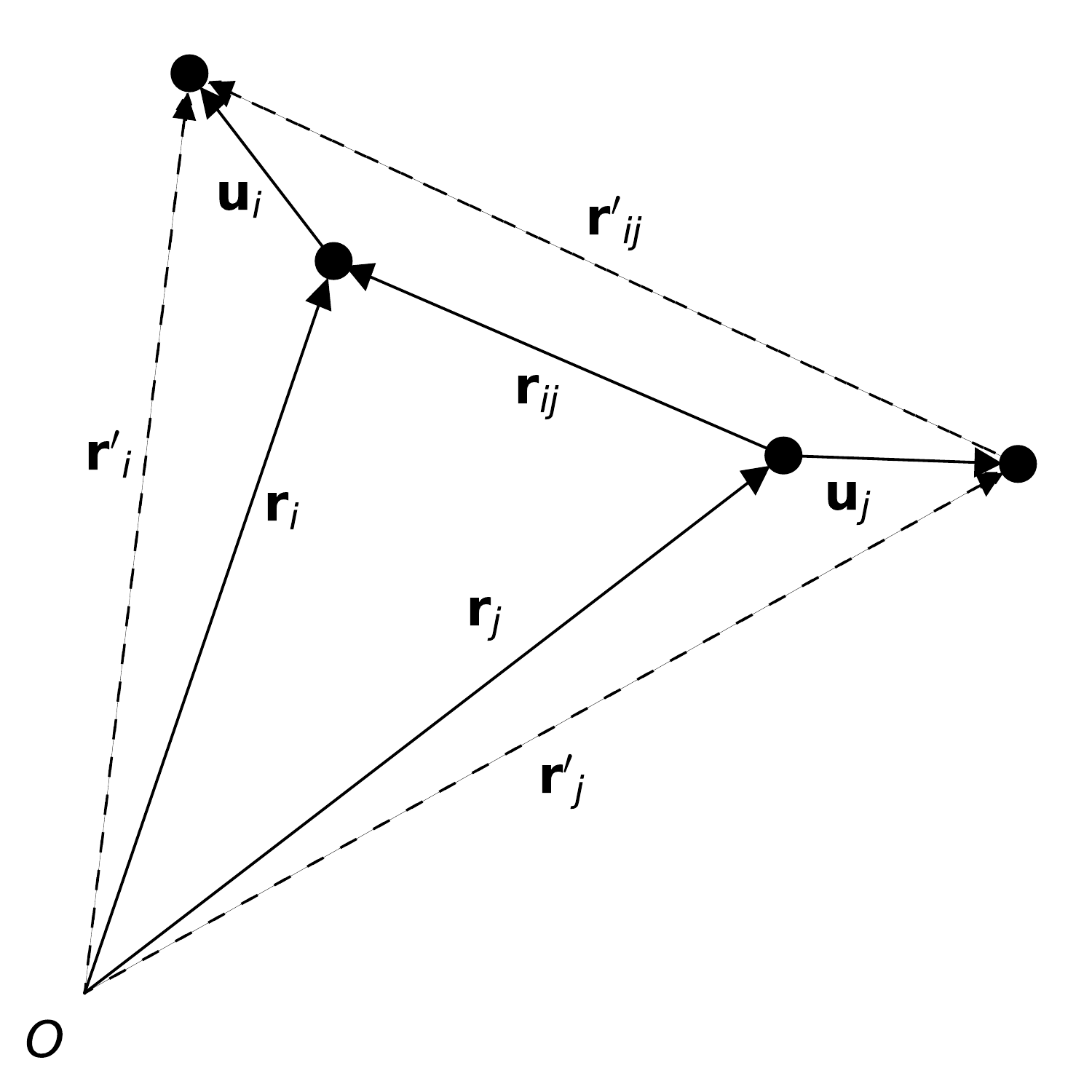}
\caption{Two interacting particles $i$ and $j$ with distance $r_{ij}$ are shown. Displacing the particles by $\mathbf{u}_{i}$ and $\mathbf{u}_{j}$ respectively, changes their separation vector $\mathbf{r}_{ij}$ to $\mathbf{r}'_{ij} = \mathbf{r}_{ij} + \mathbf{u}_{ij}$ where $\mathbf{u}_{ij} = \mathbf{u}_i - \mathbf{u}_j$.}
\label{fig:displace}
\end{figure}

Imagine $N$ particles interacting via contact potential $V(r_{ij})$ where $r_{ij} = \lvert \mathbf{r}_{i} - \mathbf{r}_{j}  \rvert$ is the distance between particles $i$ and $j$. This potential can be attractive or repulsive and can have any functional form in terms of $r_{ij}$. After applying a small deformation, particle $i$ is displaced by $\mathbf{u}_{i}$ and its new position is given by $\mathbf{r'}_{i} = \mathbf{r}_{i} + \mathbf{u}_{i}$. See Fig.~\ref{fig:displace} for reference. Total displacement between two particles, $\mathbf{u}_{ij}$, due to changes in their positions can be written as:
\begin{equation}    
  \label{eq:displacement_vector}
  \mathbf{u}_{ij} = \mathbf{u}_i - \mathbf{u}_j = \mathbf{u}_{ij,\parallel} + \mathbf{u}_{ij,\perp},
\end{equation}
where the parallel and perpendicular subscripts refer to components of the displacement vector that are parallel and perpendicular to the contact vector, $\mathbf{n}_{ij}$, connecting particle $j$ to particle $i$, defined as:
\begin{equation}
  \mathbf{n}_{ij} = \frac{\mathbf{r}_i - \mathbf{r}_j}{r_{ij}}
  = \frac{\mathbf{r}_{ij}}{r_{ij}}.
\end{equation}

\note{The distance between two particles after displacement can be written as:}
\begin{align}
    \label{eq:displacement_approx}
    r'_{ij} &= \lvert \mathbf{r'}_{i} - \mathbf{r'}_{j}  \rvert = [\left( \mathbf{r}_{ij}+\mathbf{u}_{ij} \right).\left( \mathbf{r}_{ij}+\mathbf{u}_{ij} \right)]^{1/2} \nonumber \\
    &= \left(r^{2}_{ij} + 2 {u}_{ij,\parallel} + {u}^2_{ij,\parallel} + {u}^2_{ij,\perp} \right)^{1/2} \nonumber \\
    &= \left(r_{ij} + u_{ij, \parallel}\right) \sqrt{
    1 + \left(\frac{u_{ij, \perp}}{r_{ij} + u_{ij, \parallel}}\right)^2} \nonumber \\ 
    &= \left(r_{ij} + u_{ij, \parallel}\right) \left(1 + \frac{u^2_{ij, \perp}}{2 (r_{ij} + u_{ij, \parallel}) } + \mathcal{O}(u^4) \right) \nonumber \\
    &\approx  r_{ij} + u_{ij, \parallel} + \frac{u^2_{ij, \perp}}{2r_{ij}}.
    \end{align}
\note{As a result of this approximation, the energy stored in bond $i \bondsign j$, up to second order in displacements is}:
\begin{align}
  V(r'_{ij}) &\approx V(r_{ij}) + V'(r_{ij}) (r'_{ij}-r_{ij}) + \frac{1}{2} V''(r_{ij}) (r'_{ij}-r_{ij})^2 \nonumber \\ 
  &\approx V(r_{ij}) + V'(r_{ij}) \left(u_{ij,\parallel} + \frac{u^2_{ij,\perp}}{2 r_{ij}}\right) + \frac{1}{2} V''(r_{ij}) u^2_{ij,\parallel}.
  \label{eq:energyStored}
\end{align}

In the case of harmonic potential, we can replace the second derivative of the energy with $V''(r_{ij}) = K_{ij}$ which is the stiffness of contact between particles $i$ and $j$, and the first derivative of the energy with $V'(r_{ij})= -f_{ij}$ which represents the force between two particles $i$ and $j$ due to prestress. Using this notation, we can write the change in total energy of a harmonic system as:
\begin{equation}
  \delta V=V(r') - V(r) = - \sum_{ij} f_{ij} u_{ij,\parallel} - \sum_{ij} f_{ij} \frac{u^2_{ij,\perp}}{2 r_{ij}} + \sum_{ij} \frac{1}{2} K_{ij} u^2_{ij,\parallel},
  \label{eq:energyChange_total}
\end{equation}
with $ij$ representing all of the interacting pairs of particles. The first term in Eq.~(\ref{eq:energyChange_total}) vanishes \note{when all particles are in force balance.}
Since $u^2 = u^2_{\perp} + u^2_{\parallel}$, and by defining $K'_{ij} = f_{ij}/r_{ij}$, we can write the change in total energy as:
\begin{equation}
  \delta V = \frac{1}{2} \sum_{ij} K_{ij} u^2_{ij,\parallel} - \frac{1}{2} \sum_{ij} K'_{ij} (u^2_{ij} - u^2_{ij,\parallel}).
  \label{eq:vp-v}
\end{equation}
This will be useful when we write these equations in matrix form. The parallel component of $\mathbf{u}_{ij}$ is given by $u_{ij,\parallel} = \mathbf{u}_{ij}.\mathbf{n}_{ij}$.  However, $u^2_{ij}$ cannot be written in terms of $\mathbf{n}_{ij}$. So one convenient approach is to write it as the sum of its orthogonal vector components in $d$ dimensions (\note{$\mathbf{\hat{x}_{\alpha}}$ is the unit vector along $\alpha-$axis.}):
\begin{align}
  u^2_{ij} &= u_{ij,1}^2+ u_{ij,2}^2 + ... + u_{ij,d}^2 \nonumber \\
  \newline
  &=(\mathbf{u}_{ij}.\mathbf{\hat{x}_{1}})^2 + (\mathbf{u}_{ij}.\mathbf{\hat{x}_{2}})^2 + ... + (\mathbf{u}_{ij}.\mathbf{\hat{x}_{d}})^2 .
\end{align}
Now, one can write Eq.~(\ref{eq:vp-v}) in a matrix form by defining two diagonal matrices $\mathbf{K}$ and $\mathbf{K'}$ with diagonal elements $K_{ij}$ and $K'_{ij}$, respectively. Using $\mathbf{u}^T = \left[\mathbf{u}_1,\dots,\mathbf{u}_N\right]$ as the vector of individual displacements, each term in Eq.~(\ref{eq:vp-v}) can be written as:
\begin{align}
  \label{eq:theGramTerm}
  \frac{1}{2} \sum_{ij} K_{ij} u^2_{ij,\parallel} &= \frac{1}{2} \mathbf{u}^T \mathbf{R}^T \mathbf{K} \mathbf{R} \mathbf{u} \\ \newline
  \frac{1}{2} \sum_{ij} K'_{ij} u^2_{ij,\parallel} &= \frac{1}{2} \mathbf{u}^T \mathbf{R}^T \mathbf{K'} \mathbf{R} \mathbf{u}\\ \newline
  \frac{1}{2} \sum_{ij} K'_{ij} u^2_{ij} &= \frac{1}{2} \mathbf{u}^T (\sum_{\alpha = 1}^{d} \mathbf{G}_{\alpha}^T \mathbf{K'} \mathbf{G}_{\alpha}) \mathbf{u}
  \label{eq:theGFunctions}
\end{align}
where $\mathbf{R}$ is the rigidity matrix that includes the first derivatives of constraints (bond lengths in the case of spring networks) with respect to degrees of freedom, $R_{\mu,i} = \frac{\partial r_{\mu}}{\partial x_{i}}$. $\mathbf{R}$ is a $N_c\times Nd$ dimensional matrix ($N_c$ being the number of constraints), where each column corresponds to a particle and each row, $\mu$, represents an interacting pair of particles. When grouped together, these entries make up the normalized contact vectors $\mathbf{n}_{ij}$:
\begin{equation}
\label{eq:rigidity_matrix}
  \mathbf{R} =
      \bordermatrix{& 1 & \hdots & i & \hdots & j & \hdots & N \cr
      \vdots & \vdots & \ddots & \vdots & \hdots & \vdots & \ddots & \vdots & \cr
      (i,j) & \mathbf{0} & \hdots & \mathbf{n}_{ij} & \hdots & -\mathbf{n}_{ij} & \hdots & \mathbf{0} \cr
      \vdots & \vdots & \ddots & \vdots & \hdots & \vdots &\ddots & \vdots}
\end{equation}
with $\mathbf{n}_{ji} = - \mathbf{n}_{ij}$. 
The $\mathbf{G}_\alpha$ matrix in Eq.~(\ref{eq:theGFunctions}) is given by:
\begin{equation}
  \mathbf{G}_\alpha=
  \bordermatrix{& 1 & \hdots & i & \hdots & j & \hdots & N \cr
      \vdots & \vdots & \ddots & \vdots & \hdots & \vdots & \ddots & \vdots & \cr
      (i,j) & \mathbf{0} & \hdots & \mathbf{\hat{x}_{\alpha}} & \hdots & -\mathbf{\hat{x}_{\alpha}} & \hdots & \mathbf{0} \cr
      \vdots & \vdots & \ddots & \vdots & \hdots & \vdots &\ddots & \vdots}
\end{equation}
\newline
which is also a $N_c\times Nd$ dimensional matrix with unit vectors $\mathbf{\hat{x}_{\alpha}}$ in each orthogonal direction. For instance, if $\alpha = 1$, then $\mathbf{G}_{1}$ for a $3D$ system will contain $\mathbf{\hat{x}_{1}} = [1, 0, 0]$ vectors only.

Using Eqs.~(\ref{eq:theGramTerm}-\ref{eq:theGFunctions}), we can write the change in total energy as:
\begin{align}\label{eq:energy_change_matrix}
    \delta V = \frac{1}{2} \mathbf{u}^T \overbrace{
    \left(
    \underbrace{\mathbf{R}^T \mathbf{K} \mathbf{R}}_\text{$\mathbf{H}_{\text{g}}$}
    + \underbrace{\mathbf{R}^T \mathbf{K'} \mathbf{R} - (\sum_{\alpha = 1}^{d} \mathbf{G}_{\alpha}^T \mathbf{K'} \mathbf{G}_{\alpha})}_\text{$\mathbf{H}_{\text{p}}$} 
    \right)
    }^\text{$\mathbf{H}$} \mathbf{u}
\end{align}
\note{where $\mathbf{H}_{\text{g}}$ and $\mathbf{H}_{\text{p}}$ are the geometrical and prestress terms in the Hessian, $\mathbf{H}$, respectively. When a system has no prestress forces, the Hessian reduces to the geometrical part only}:
\begin{equation}
    \label{eq:full_hessian_no_prestress}
   \mathbf{H}_{\text{no-prestress}} = \mathbf{H}_{\text{g}} = \mathbf{R^T K R}.
\end{equation}
This form of the Hessian is also called the dynamical matrix.

Note that entries of $ \mathbf{H}$ match the mathematical definition of Hessian which includes second derivatives of the energy with respect to degrees of freedom. This means that the Hessian in Eq.~(\ref{eq:energy_change_matrix}) can also be derived directly using its definition:
 \begin{equation}
   \label{eq:full_hessian_derivative}
   \mathbf{H}_{ij}^{\alpha \beta} = \frac{\partial^2 V}{\partial r_i^{\alpha} \partial r_j^{\beta}} \ . 
 \end{equation}
In the case of a harmonic system, where the energy is given by:

\begin{equation}
    \label{eq:harmonic_energy}
   V = \frac{1}{2} \sum_{i,j} K_{ij} \ ({r'}_{ij} - {r}_{ij})^2,
\end{equation}
writing all the derivatives in Eq.~(\ref{eq:full_hessian_derivative}) leads to terms that are linear in $({r'}_{ij} - {r}_{ij})$ and terms that are independent of $({r'}_{ij} - {r}_{ij})$. Those terms that are linear in $({r'}_{ij} - {r}_{ij})$ represent the forces between pairs of interacting particles ($f_{ij}$) and thereby belong to the $\mathbf{H}_{p}$ matrix. On the other hand, terms that are independent of the changes in distance between two particles, represent the geometrical Hessian, $\mathbf{H}_{g}$.  \note{Harmonic potentials are widely used in models of elastic networks such as Anisotropic Network Model~\cite{atilgan2001anisotropy,doruker2000dynamics} and Gaussian Network Model~\cite{haliloglu1997gaussian,bahar1997direct}. There are a variety of existing software packages for molecular rigidity analysis that make use of these models~\cite{bakan2011prody,li2016gnm}. However, we note that the formalism we present here is distinct as it is extendable to any central-force energy function. In addition, \rp{} provides implementations of various boundary conditions as well as a set of crucial tools such as the rigidity matrix, the geometrical and prestress terms of the Hessian, and elastic moduli, which make it suitable for analysing the rigidity and flexibility of a larger class of elastic systems including higher-order rigid networks~\cite{damavandi2021energetic}.} 


\subsection{Zero Modes and Infinitesimal Zero Modes}
The vibrational modes of a system are the eigenmodes of the Hessian matrix given in Eq.~(\ref{eq:energy_change_matrix}). Number of zero modes, thereby, represents the number of ways in which one can perturb the system without any change in the energy~\note{\cite{hinsen2005normal}}. 
Infinitesimal zero modes are the zero modes of the geometrical part of Hessian which is equivalent to assuming that there is no prestress in the system:
\begin{align*}
    \label{eq:infinitesimal}
    & \text{Infinitesimal zero modes} = \\
    \newline
    & \text{Zero modes of the geometrical Hessian (dynamical matrix),} \ \mathbf{ R^T K R} = \\
    \newline
    & \text{Zero modes of a system with no prestress} 
\end{align*}
Note that when infinitesimal zero modes and overall zero modes in a system only include trivial rigid motions (translations and rotations), the system is first-order rigid\note{~\cite{connelly1993rigidity}}. On the other hand, when the geometrical part of the Hessian has non-trivial zero modes (e.g. when the system is under-constrained), but the overall Hessian only has trivial zero modes, the system is said to be second-order rigid\note{~\cite{connelly1993rigidity, damavandi2021energetic2}}. In this case, the prestress Hessian is positive definite and its eigenvalues can balance the non-trivial zero modes of the geometrical Hessian, leading to second-order rigidity in the system~\cite{damavandi2021energetic,damavandi2021energetic2}.

\subsection{States of Self-Stress}
States of self-stress refer to possible ways one can put non-zero forces on contacts while keeping the system at mechanical equilibrium with zero resultant force on each particle or node~\cite{lubensky2015phonons}. In a system of particles, the total force on each particle is the sum of all the forces exerted by its interacting neighbors. The total force on particle $i$ is related to the contact forces shared between $i$ and its interacting neighbors through the equilibrium matrix, which is the transpose of the rigidity matrix:

\begin{equation}
    \label{eq:forces_on_particles}
   \mathbf{F}_i = \mathbf{R^T}_{i \ell} \ \mathbf{\tau}_{\ell},
\end{equation}
where $\mathbf{\tau}_{\ell}$ is the force on the $\ell$th neighboring contact. States of self-stress, therefore, are all the non-trivial solutions to $\mathbf{R^T}\mathbf{\tau} = 0$ which gives the right null-space of the equilibrium matrix $\mathbf{R^T}$\note{~\cite{pellegrino1993structural}}. To find the right null-space of this matrix, one can multiply $\mathbf{R^T}\mathbf{\tau} = 0$ by the rigidity matrix, $\mathbf{R}\mathbf{R^T}\mathbf{\tau} = 0$, and find zero eigenvalues of the resulting matrix, $\mathbf{N}$:

\begin{equation}
    \label{eq:N_matrix}
   \mathbf{N} = \mathbf{RR^T}, 
\end{equation}
In other words:
\begin{align*}
    \label{eq:sss}
    & \text{States of self-stress} = \text{Zero modes of matrix} \ \mathbf{N}.
\end{align*}

\subsection{Elastic Properties}
The elasticity of a system is measured by its response to an applied deformation. When a system is first-order rigid, elastic moduli of the system (shear modulus in particular) can be used to determine its rigidity~\cite{damavandi2021energetic}. 
According to Hooke's law, the stress induced in an elastic material is proportional to the amount of strain (deformation) that has been applied to it. In the linear regime (harmonic approximation), the energy density ($\delta v = \delta V / \text{Volume}$) stored in an elastic object due to a strain can be written as:
\begin{equation}\label{eq:energydensity}
\delta v = \frac{1}{2} \sum_{ij} c_{ij} e_{i} e_{j},
\end{equation}
where $e_i$ is the strain in direction $i$ and $c_{ij}$ are the moduli of elasticity, characterizing the resistance of the material to elastic deformations~\cite{hagh2018rigidity}. As an example, in $2D$, the strain matrix is written as:
\begin{equation}
\label{eq:strainmatrix}
    \begin{pmatrix}
        e_{xx} & e_{xy}  \\
        e_{xy} & e_{yy} 
    \end{pmatrix}.
\end{equation}

Elastic moduli can be calculated by determining the direction and magnitude of the applied deformation. Bulk modulus and shear modulus are two of the primary moduli that are widely used in the study of elastic response in solids. Bulk modulus, $B$, measures the response of a system to a uniform compression in all directions. In a $2D$ system, the strain matrix for uniform compression can be written as:
\begin{equation}
    \begin{pmatrix}
        -\epsilon & 0  \\
        0 & -\epsilon
    \end{pmatrix},
\end{equation}
which leads to
\begin{equation}
    \delta v = \frac{1}{2}
\left(c_{11} + c_{11} + 2 c_{12}\right) \epsilon^2 = 2 B \epsilon^2.
\end{equation}
Shear modulus, on the other hand, is a measure of the material's response when it experiences a force parallel to one of its surfaces while the opposite surface undergoes a deformation in the opposite direction. There are two common types of shear in $2D$ with the following strain matrices:

\begin{equation}
    \begin{pmatrix}
        \epsilon & 0  \\
        0 & -\epsilon
    \end{pmatrix},
    \begin{pmatrix}
        0 & \epsilon  \\
        \epsilon & 0
    \end{pmatrix},
\end{equation}
which are known as the \textit{pure shear}, $G$, and \textit{simple shear}, $G_{xy}$, respectively. Note that in both shear cases, the volume of the material is preserved. The corresponding shear moduli are found to be:
\begin{align}
    \delta v &= \frac{1}{2}
\left(c_{11} + c_{11} - c_{12}\right) \epsilon^2 = 2 G \epsilon^2 \\
\delta v &= \frac{1}{2}
\left(c_{33} \right) \epsilon^2 = 2 G_{xy} \epsilon^2.
\end{align}
For a more detailed review of elastic moduli and their derivations in $2D$, see~\cite{hagh2018rigidity}.

\section{Program Description and Installation}

In this paper, we introduce \rp{} which is a lightweight Python library dedicated to studying rigidity and linear response in spring networks. The flexibility of \rp{} allows the user to easily combine its outputs with other scientific tools in Python.
\rp{} uses an object-oriented style of programming which gives the user access to a set of methods to compute the desired quantities or extend the library with custom functions. This functionality is made possible by expressing the network information and its dynamics as a set of linear equations that relate the changes in constraints to changes in degrees of freedom (See Section~\ref{sec:theory}). \rp{} has been successfully used in multiple research projects~\cite{sadjadi2021realizations, sadjadi2018two, hagh2019broader} and its latest version is accessible on GitHub~\cite{rpgit}.

\rp{} is written with Python $3$ in mind. However, it is also tested successfully in Python $2.7$, although this version is no longer supported. The package is written so that the dependencies are limited to the standard scientific libraries of Python such as Numpy and Scipy~\cite{scipy}. The code takes full advantage of the vectorization and fast performance of the BLAS/LAPACK library. However, for those with Intel CPUs,  it is recommended to build the numpy/scipy environments with Intel(R) Math Kernel Library~\cite{intel} for optimal performance. 

The most convenient way to install the package is through Python Package Index~\cite{pypi} and \texttt{pip} package manager:
\begin{verbatim}
>>> pip install rigidpy 
\end{verbatim}
However, if the user prefers to build from the source or install the development version, the package can be downloaded from GitHub~\cite{rpgit} by:
\begin{verbatim} 
>>> git clone https://github.com/vardahagh/rigidpy.git
\end{verbatim}
To install the package, change directory to \rp{} and use the following command:
\begin{verbatim}
>>> pip install --no-cache-dir .
\end{verbatim}
To improve the package usability, we have divided the application programming interface (API) into several modules based on their functionalities. Currently, \rp{} consists of three modules:
\begin{itemize}
    \item \texttt{framework} module is the base class. It receives lists of node positions, bonds, boundary conditions, stiffnesses, masses, and rest lengths to construct a \texttt{framework} object. This class provides multiple methods to compute the rigidity matrix, geometrical and prestress Hessian matrices, states of self-stress, elastic moduli, \textit{etc}.
    \item \texttt{configuration} provides functionality for geometry optimization and energy minimization. \rp{} currently supports Newton-Conjugate Gradient and L-BFGS-B optimization algorithms. The L-BFGS-B algorithm can be used in optimizations that fix the positions of a subset of nodes.  
    \item \texttt{circuit} contains two methods to find alternative realizations of a network. The methods are based on constraint reduction and cell-volume change. For more information, see~\cite{sadjadi2021realizations}.
\end{itemize}

In the following section, we show a detailed example of applying \rp{} to a periodic spring network using the \texttt{framework} module.  One can find further information about \texttt{configuration} and \texttt{circuit} modules in the \rp{} repository.

\section{Usage}
At the core of \rp{} there is \texttt{framework} class that creates a base class to compute the rigidity and elastic properties of a network. 
To make a \texttt{framework} object, a minimum of two arrays/lists are required: (i) coordinates of nodes, (ii) list of bonds. Users can save the coordinates and bond list in any desired format, but the inputs to \texttt{framework} have special shape requirements to ensure the correct rigidity characterization. For a network with $N$ nodes in $d$ dimensions and $N_c$ connecting bonds, the shape of ``coordinates'' array is $(N, d)$ and of the ``bonds'' array is $(N_c,2)$. The inputs can be Numpy arrays or Python lists. 

\begin{figure}
    \centering
    \includegraphics[scale=0.4]{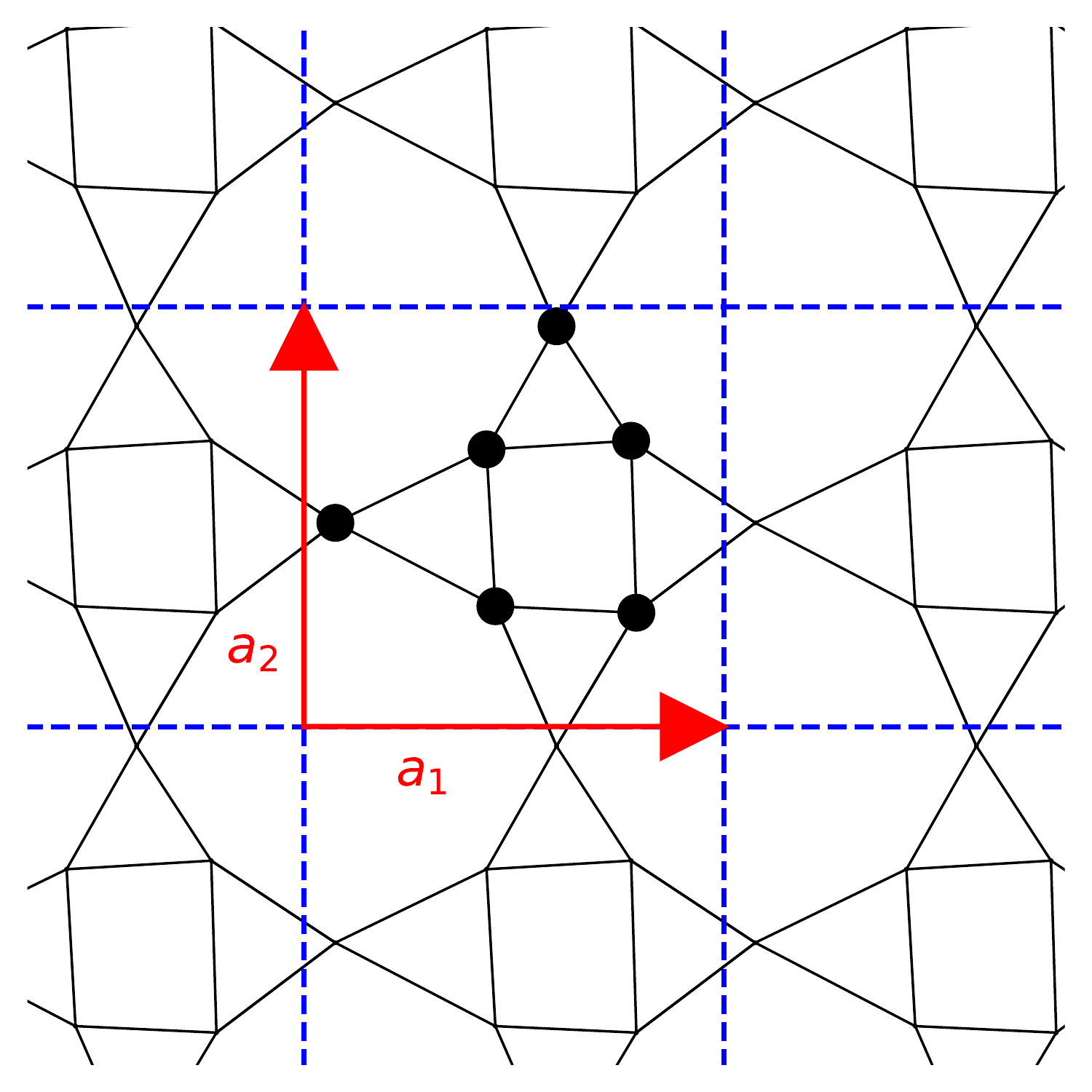}
    \caption{A $2D$ example of a periodic network with $6$ nodes and $12$ bonds in the unit cell. The red arrows show the repeat/basis vectors.}
    \label{fig:framework}
\end{figure}

As was discussed earlier, boundary conditions can greatly impact the rigidity of a network. If no boundary conditions are specified, \texttt{framework} defaults to \textit{free} boundaries which is equivalent to having no constraints on the positions of the nodes. However, \rp{} also supports periodic and anchored (pinned) boundary conditions~\cite{theran2015anchored} by setting the \texttt{basis} and \texttt{pins} parameters. 
The \texttt{basis} parameter is used to specify the array of repeat vectors with shape $(d,d)$. The \texttt{pins} parameter, receives a list of node indices and fixes their positions by effectively freezing the relevant degrees of freedom~\cite{theran2015anchored}. Note that \texttt{basis} takes precedence over \texttt{pins} parameter. This enables users to create periodic lattices while pinning a selection of nodes. If ``pure anchored boundary conditions'' are desired~\footnote{Pure anchored boundary conditions are satisfied when exactly half of the nodes on the surface of a network are immobilized, but users can provide fewer or more pins.}, users should not provide any \texttt{basis} vectors and only provide a list of nodes to pin. 

Fig.~\ref{fig:framework} shows a $2D$ network with periodic boundary conditions. The coordinates and bonds of the network can be found in the GitHub repository under \texttt{tests/data\_6} directory. The coordinates are saved in a file named \texttt{coordinates.txt} where each row represents a coordinate pair $(x,y)$:
\begin{verbatim}
  1.398598   2.732305
  0.964475   1.940684
  0.072057   1.408286
  0.875796   0.956616
  1.891248   0.873234
  1.951186   1.936490
\end{verbatim}
The nodes are indexed in the order they appear inside the coordinates file. The list of bonds is provided in \texttt{bonds.txt} file, where each row represents a pair of connected nodes $i \bondsign j$:
\begin{verbatim}
   0    1
   0    3
   0    4
   0    5
   1    5
   1    2
   1    3
   2    3
   2    4
   2    5
   3    4
   4    5
\end{verbatim}
Note that each bond appears only once (since $1 \bondsign 0$ is the same bond as $0 \bondsign 1$) and the node indices start at zero. 
The basis vectors for the network in Fig.~\ref{fig:framework} are saved in \texttt{basis.txt}:
\begin{verbatim}
  2.732051   0.000000
  0.000000   2.732051
\end{verbatim}
Coordinates, bonds, and basis vectors can be imported via Numpy. Once \rp{} is imported, a \texttt{framework} can be easily constructed~\footnote{Users can find this example under \url{https://github.com/VardaHagh/Rigidpy/tree/master/notebooks/basic_example.ipynb} in the GitHub repository with additional information.}:
\begin{verbatim}
    >>> import rigidpy as rp
    >>> import numpy as np
    >>> coordinates = np.loadtxt("./tests/data_6/coordinates.txt")
    >>> bonds =  np.loadtxt("./tests/data_6/bonds.txt",int)
    >>> basis = np.loadtxt("./tests/data_6/basis.txt")
    >>> restLengths = 1.0
    >>> F = rp.framework(coordinates, bonds, basis, restLengths=restLengths)
\end{verbatim}
\note{In the example above, \texttt{framework} is created using the required arguments (coordinates, bonds, and basis) plus the optional argument, \texttt{restLengths}, which is necessary for networks with prestress.
Users can specify other network properties such as spring constants (\texttt{k}), list of pinned particles (\texttt{pins}), potential power (\texttt{power}), \textit{etc}. Note that both the spring constants and rest lengths can be specified for each bond separately}. 
\note{If the network is not in mechanical equilibrium, it should be relaxed \textit{before} the \texttt{framework} is created. This can be achieved using the \texttt{configuration} module:}
\begin{verbatim}
    >>> config = rp.configuration(coordinates, bonds, basis)
    >>> relaxedCoordinates = config.energyMinimizeNewton(restLengths, restLengths)
    >>> F = rp.framework(relaxedCoordinates, bonds, basis, restLengths=restLengths)
\end{verbatim}
\note{Note that in the definition of \texttt{F}, we have used \texttt{relaxedCoordinates} instead of \texttt{coordinates}, since these are the coordinates of the nodes after the system has been brought to a local energy minimum.}
Once \texttt{F} is defined as a \texttt{framework} object, the user has access to a set of functions to compute the rigidity matrix, the Hessians, and the elastic moduli of the network which are presented in this section. \note{In addition, one can produce a simple visualization (as seen in Fig. \ref{fig:framework}) for networks with free, periodic, and pinned boundaries using the \texttt{visualize} function:}
\begin{verbatim}
    >>> F.visualize()
\end{verbatim}

\subsection{Rigidity Tools}
One could easily compute the rigidity matrix, geometrical Hessian, prestress Hessian, and full Hessian matrices using the \texttt{framework} object:
\begin{verbatim}
    >>> R = F.rigidityMatrix()
    >>> hessianMatrixGeometric = F.hessianMatrixGeometric()
    >>> hessianMatrixPrestress = F.hessianMatrixPrestress() 
    >>> fullHessianMatrix = F.hessianMatrix()
\end{verbatim}
\note{Note that to detect any prestress forces in the network, one should provide the rest length values when creating the \texttt{framework} object. Otherwise, \rp{} assumes that the rest lengths are the same as the bond lengths, giving zero values for the prestress term in the Hessian.} 

Users can also compute the eigenvalues and eigenvectors of the Hessian matrix. For instance, to compute the first $5$ smallest eigenvalues and their corresponding eigenvectors of the full Hessian, one can use:
\begin{verbatim}
    >>> eigenValues, eigenVectors = F.eigenSpace(eigvals=(0, 4))
\end{verbatim}
To compute the entire set of eigenvalues simply pass \texttt{eigvals=None}. The results are floating-point numbers and whether an eigenvalue is zero is the user's choice. Zero modes can be found by setting a threshold on the eigenvalues of the Hessian. Similarly, infinitesimal zero modes can be found by computing the eigenvalues of the \verb!hessianMatrixGeometric!. Finally, states of self-stress are computed by:
\begin{verbatim}
    >>> SSS = F.selfStress()
    >>> print (SSS.shape)
    (12,2)
\end{verbatim}
In this particular example, there are two states of self-stress since the network has two bonds in excess of isostaticity. 

\subsection{Elastic moduli}
\rp{} has built-in functions to compute the bulk and \textit{pure} shear moduli of a network. Both these moduli can be computed using the \texttt{framework} object:
\begin{verbatim}
    >>> B = F.bulkModulus()
    >>> G = F.shearModulus()
    >>> print ("bulk modulus = {:.2f}, shear modulus = {:.2f}".format(B, G))
    bulk modulus = 0.35, shear modulus = 0.18
\end{verbatim}
To measure the \textit{simple} shear, one needs to provide the appropriate strain matrix (see the following example).
In general, the \texttt{elasticModulus} function within \texttt{framework} can measure an elastic modulus for any arbitrary deformation within linear regime. The strain (deformation) matrix is of the shape $(d,d)$ as given in Eq.~(\ref{eq:strainmatrix}), and the energy stored in the network due to a given deformation is computed using Eq.~(\ref{eq:energydensity}). Note that the strain matrix passed to the function is the strain matrix plus the identity matrix in $d$ dimensions.

As an example, consider a deformation consisting of an elongation along $y-$axis and a contraction along $x-$axis but with no changes along $xy$ plane. The strain matrix can be defined as:
\begin{verbatim}
    >>> eps = 1e-6
    >>> strainMatrix = np.array([[1-eps, 1],[1,1+eps]])
    >>> print (strainMatrix)
    [[0.999999 1.0]
     [1.0      1.000001]]
\end{verbatim}
To compute the elastic modulus after applying this deformation on the network in Fig.~\ref{fig:framework}, one can simply use: 
\begin{verbatim}
    >>> em = F.elasticModulus(strainMatrix)
    >>> print (em)
    0.011521283842869033
\end{verbatim}
\section{Conclusion and Future Work}

In this paper, we provide an overview of the mathematical framework within which the rigidity and mechanical response of \note{central-force} systems can be studied. We then introduce a new Python library, \rp{}, that includes the tools and modules necessary for computing linear response in spring networks, in addition to their elastic moduli and vibrational modes. The authors would like to continue the development process with the support of the community by adding new tools and features to the library. 

Rigidity research is not limited to Hookean spring networks with central-force potentials. There are many materials and systems in which the bending energy must be included for more realistic modeling of the system~\cite{overney1993structural, kang2013molecular,rens2019rigidity}. In other cases, non-Hookean energy functions such as Hertzian potentials and density-independent models such as Vertex Model are used~\cite{o2003jamming, corwin2005structural,bi2015density}. One exciting addition to \rp{} would be to include these models by adding automatic differentiation tools that can automatically differentiate arbitrary energy functions without writing explicit equations that relate the constraints to degrees of freedom~\cite{jax2018github, jaxmd2020}. 

A technical limitation of the current implementation is that matrices are represented as dense arrays which limit \rp{}'s scalability (we have successfully tested our implementation with system sizes of up to $\mathcal{O}$($10^3-10^4$) particles). However, for short-range energy functions, both the Hessian and rigidity matrices are very sparse. The support for sparse representation is another feature that the authors would like to add to \rp{}, in the hope that it will make the package more suitable for larger system sizes.

\section{Acknowledgements}
This work has been supported by National Science Foundation under grant DMS 1564468 (MS and VFH) and by Simons Foundation via award 348126 (VFH). 

\bibliography{rigidPyPaper.bib}
\pagebreak

\end{document}